\documentclass[11pt]{article}

\usepackage[normalem]{ulem}
\usepackage{longtable}
\usepackage{makeidx}
\usepackage{fullpage}
\usepackage{amsmath}
\usepackage{amssymb}
\usepackage{setspace}
\usepackage{graphics}
\usepackage{epsfig}
\usepackage[font=footnotesize,labelfont=bf,justification=centerlast,width=.94\textwidth]{caption}
\usepackage{cite}
 \usepackage{multirow}
\usepackage{array,booktabs}
\usepackage[cal=boondox]{mathalfa}
\usepackage{xcolor}

\usepackage{hyperref}

\hypersetup{
    bookmarks=true,%
    colorlinks,%
    citecolor=blue,%
    filecolor=black,%
    linkcolor=black,%
    urlcolor=blue
}

\newcommand{\be}{\begin{equation}}
\newcommand{\ee}{\end{equation}}
\newcommand{\bes}{\begin{equation*}}
\newcommand{\ees}{\end{equation*}}
\newcommand{\bea}{\begin{eqnarray}}
\newcommand{\eea}{\end{eqnarray}}
\newcommand{\beas}{\begin{eqnarray*}}
\newcommand{\eeas}{\end{eqnarray*}}

\newcommand{\bmat}{\begin{bmatrix}}
\newcommand{\emat}{\end{bmatrix}}

\newcommand{\slt}{SL(2,\RR)}

\newcommand{\RR}{\mathbb{R}}

\def\tr{{\rm tr}}
\def\Tr{{\rm Tr}}
\def\le{\left}
\def\ri{\right}

\def\Tr{{\rm Tr}}
\def\le{\left}
\def\ri{\right}

\newcommand{\ben}{\begin{enumerate}}
\newcommand{\een}{\end{enumerate}}

\begin{document}
\numberwithin{equation}{section}
{
\begin{titlepage}
\begin{center}

\hfill \\
\hfill \\
\vskip 0.75in

{\Large \bf General solutions in Chern-Simons gravity  and $T \bar T$-deformations }\\

\vskip 0.4in

{\large  Eva Llabr\'es${}$}\\

\vskip 0.3in

{\it Universit\'e Paris-Saclay, CNRS, CEA, Institut de physique th\'eorique, 91191, Gif-sur-Yvette, France} \vskip .5mm

 \vskip .5mm

\texttt{eva.llabres-llambias@ipht.fr}

\end{center}

\vskip 0.35in

\begin{center} {\bf ABSTRACT } \end{center}

We find the most general solution to Chern-Simons AdS$_3$ gravity in Fefferman-Graham gauge. The connections are equivalent to geometries that have a non-trivial curved boundary, characterized by a 2-dimensional vielbein and a spin connection. We define a variational principle for Dirichlet boundary conditions and find the boundary stress tensor in the Chern-Simons formalism. Using this variational principle as the departure point, we show how to treat other choices of boundary conditions in this formalism, such as, including the mixed boundary conditions corresponding to a $T \bar T$-deformation. 

%This formalism allows us to propose a boundary term for the Chern-Simons action which is equivalent to perform a $T \bar T$-deformation in the boundary theory.   

\vfill

\noindent \today

\end{titlepage}
}
%%%%%%%%%%%%%%%%%%%%%%%%%%%%%%%%%%%%%%%%%%%%%%%%%%%%%%%%%%%%%%%%%%

%%%%%%%%%%%%%%%%%%%%%
\newpage

\tableofcontents

\section{Introduction}

Three-dimensional gravity is a topological theory with no local degrees of freedom. This feature makes it a useful framework to study aspects of quantum and classical gravity, avoiding some of the complications we encounter in higher dimensions. The topological character of  AdS$_3$ gravity is made manifest by rewriting it as a Chern-Simons theory with gauge group $SO(2,2) = \slt \times \slt$ \cite{Achucarro:1987vz,Witten:1988hc}. The Chern-Simons formulation has many advantages: the Einstein field equations map to a flatness condition on the connection, diffeomorphisms may be easily interpreted as gauge transformations \cite{Witten:1988hc}, BTZ black holes emerge naturally as topological defects around which the gauge fields have non-trivial holonomies.  Moreover, the Chern-Simons formulation simplifies in a great manner the study of theories of higher spin gravity \cite{Aragone:1983sz,Blencowe:1988gj,Bergshoeff:1989ns,Henneaux:2010xg,Campoleoni:2010zq}. In this paper, we study 3d gravity in Chern-Simons formalism. This theory should not be confused with Einstein gravity plus a Chern-Simons term since both theories are sometimes referred to as Chern-Simons gravity.

Since there are no propagating degrees of freedom in the bulk, boundary conditions are especially relevant, because they encode the interesting physics. Boundary conditions have been extensively studied in the metric formulation \cite{Brown:1986nw,Brown:1992br,Henningson:1998gx,Balasubramanian:1999re}.  The analysis is highly simplified by the fact that the most general asymptotically AdS$_3$ condition is known
\be\label{FG0}
ds^2=\ell^2{dz^2\over z^2} +g_{ij}(x^k, z)dx^idx^j,     \qquad\qquad g_{ij}(x^k,z)={g_{ij}^{(0)}(x^k)  \over z^{2}}+g_{ij}^{(2)}(x^k)+ z^{2}\,g_{ij}^{(4)}(x^k)\,,
\ee 
where $z$ is a radial coordinate, with boundary at $z\rightarrow 0$, and $i,\,j,\,k$ are 2-dimensional boundary indexes. Equation \eqref{FG0} is known as the Fefferman-Graham expansion of the metric. The parameters $g_{ij}^{(2)}$, and $g_{ij}^{(4)}$ are related to $g_{ij}^{(0)}$ through the asymptotic equations of motion \cite{Skenderis:1999nb}. Different boundary conditions correspond to fixing different components of the asymptotic metric. For example, $g_{ij}^{(0)}$ can be fixed with Dirichlet boundary conditions, we could impose Neumann for $g_{ij}^{(2)}$, or consider a situation with mixed boundary conditions. These have a clear interpretation in the context of the AdS$_3$/CFT$_2$ correspondence: when $g_{ij}^{(0)}$ is fixed we can identify it with the CFT$_2$ metric \cite{Balasubramanian:1999re}, Neumann boundary conditions couple the boundary to 2d gravity, and mixed conditions are equivalent to double-trace deformations in CFT \cite{Klebanov:1999tb,Witten:2001ua}. However, a systematic analysis of the Fefferman-Graham expansion \eqref{FG0} in Chern-Simons formalism is missing in the literature. This could be very helpful since the gauge theory description of 3d gravity is generally simpler than the metric case. In this paper, we solve the Chern-Simons equations of motion in a systematic way for a set of boundary conditions that reproduce the Fefferman-Graham gauge \eqref{FG0}. Generalized boundary conditions for Chern-Simons AdS$_3$  gravity were already discussed in \cite{Grumiller:2016pqb,Cotler:2018zff}. In Section \ref{compmet}, we will comment on how our generalized connections relate to the results in \cite{Grumiller:2016pqb,Cotler:2018zff}.

Double-trace deformations in CFT have recently obtained much attention in the context of holography.  This is mainly due to the proposal  \cite{McGough:2016lol} which relates $T\bar T$-deformed CFT$_2$ to AdS$_3$ gravity with a finite bulk cutoff. This proposal was generalized to include matter in \cite{Kraus:2018xrn},  higher dimensions \cite{Taylor:2018xcy,Hartman:2018tkw,Caputa:2019pam}, and entanglement entropy calculations \cite{Chen:2018eqk,Murdia:2019fax,Ota:2019yfe,Banerjee:2019ewu}. 
%Single-trace deformations in the boundary of Chern-Simons gauge theories were previously considered in the context of higher spin black holes \cite{deBoer:2014fra}. 
However, the formulation of double-trace deformations in AdS$_3$ in Chern-Simons formalism has remained unstudied until now. We will use our new generalized connections to define a variational principle with Dirichlet boundary conditions as in the metric formalism. This allows us to interpret the boundary values of the gauge fields as sources and expectation values of the dual operators in the CFT. This identification will prove very useful because it constitutes a departure point for the analysis of double-trace deformations in Chern-Simons formalism. We focus on one particular type of deformation that will be dual to a $T\bar T$-deformation in the boundary CFT. 
%Higher spin fields have a very natural generalization in Chern-Simons formalism, and our analysis of this specific deformation could lead to an efficient way of treating general types of double-trace deformations.  
We would like to emphasize that our analysis is purely classical. The identification between the boundary terms of the Chern-Simons theory and the deformed CFT action is done using known results from the metric formalism \cite{Guica:2019nzm}. The precise understanding of the subtleties that might arise at the quantum level is left for future work.
 
This article is organized as follows. We start in Section \ref{sec:CS} by reviewing the Chern-Simons formalism of 3d gravity. In Section \ref{newsol}, we solve the $SL(2,\RR)\times SL(2,\RR)$ Chern-Simons theory for a set of generalized boundary conditions, and construct a well-defined variational principle. In Section \ref{compmet},  we identify our connections with the most general solutions of asymptotically AdS$_3$ spaces in the Fefferman-Graham gauge and give a holographic interpretation to the boundary values of the connections. Finally, in Section \ref{sec:DTdef}, we use the knowledge acquired in the previous sections to analyze double-trace deformations in the Chern-Simons formalism. We conclude in Section \ref{sec:con}, and in the Appendices, we set our notation and review the vielbein formalism of general relativity, which is useful for our calculations. 

\section{Review: AdS$_3$ as a Chern-Simons theory}\label{sec:CS}

There exists an alternative description of pure 3d gravity in terms of Chern-Simons (CS) gauge connections. In this section, we briefly review this formalism, and for further details, we refer the reader to the original articles \cite{Achucarro:1987vz,Witten:1988hc} and more recently to, e.g., \cite{Banados:1998gg,Ammon:2012wc}.  Let us start by considering the Chern-Simons action:
\bea\label{actiona}
S_{CS}[{A}] =\frac{k}{4\pi}\int_{\cal M} \Tr\left({ A} \wedge d{A} + \frac{2}{3} {A} \wedge {A} \wedge { A}\right)~,
\eea
where $A$ is a gauge field that lives in a manifold ${\cal M}$. The trace $\text{Tr}(...)$ is a shortcut notation for the contraction using the Killing forms of the algebra. The gauge group is chosen to match the symmetry group of AdS$_3$, which in Lorentzian signature is  $SO(2,2)\cong SL(2,\RR)\times SL(2,\RR)$. As a result, we consider the gauge connections ${A,\,\bar A}$ valued in two independent copies of $sl(2,\mathbb{R})$, which are defined as %\textcolor{blue}{(Maybe write why CS is good when I have the intro)}:
\be\label{vbspin}
A= (\Omega^m +{1\over \ell} E^m)L_m ~,\quad  \bar A= (\Omega^m -{1\over \ell}E^m)\bar L_m\,,
\ee
where $E^m$, and $\Omega^m$, and $L_m$ are respectively the 3-dimensional vielbein, spin connection, and $sl(2,\mathbb{R})$ generators defined in Section \ref{app:sl2}. With the definition in \eqref{vbspin}, the following actions are equivalent: 
\bea\label{actiona}
S_{\rm EH} [E,\Omega]= S_{CS}[A] - S_{CS}[\bar A]~,
\eea
where  $S_{\rm EH}$ is the Einstein-Hilbert action with negative cosmological constant, written in vielbein formalism. The relation \eqref{actiona} is only true if the gravitational constant is related to the Chern-Simons level via
  \be\label{eq:level}
  k={\ell\over 4G_3}~,
  \ee
%
%The advantage of the Chern-Simons formalism of 3d gravity is that in most situations the gauge theory is easier to manipulate . For example, the Einstein field equations are reduced to the flatness condition on the gauge connection:
Moreover, one can show that the Einstein field equations are equivalent the Chern-Simons equations of motion
\be\label{eom}
d{A} + {A} \wedge { A}=0\,,\qquad\qquad d{\bar A} + {\bar A} \wedge { \bar A}=0\,,
\ee
The space-time metric can be recovered directly from the connections through
\be\label{metsl2}
g_{\mu\nu} =2 \Tr (E_{\mu} E_{\nu})={1\over 2}\Tr \left((A-\bar A)_{\mu} (A-\bar A)_{\nu}\right)~,
\ee
where we are taking the trace in the fundamental representation of $sl(2,\mathbb{R})$, with the conventions in \eqref{sl2metric}.  This provides a map between the solutions of the Chern-Simons theory with $SL(2,\RR)\times SL(2,\RR)$ and geometries in AdS$_3$.

The construction of locally AdS$_3$ solutions in the CS formalism s discussed in \cite{Banados:1998gg}, whose results we summarize here. We start by gauge fixing the radial component of the connection to
\begin{align}\label{radial}
A_{\rho}\,=\,L_0\,,\qquad \qquad  \bar A_{\rho}\,=\,-L_0\,\,.
\end{align}
With this gauge choice, the connections can be parametrized without losing any generality \cite{Campoleoni:2010zq} by
\begin{align}\label{eq:aba}
A= b(\rho)^{-1}\le(a(x^+,x^-) + d\,\ri)b(\rho)\,,\qquad \bar{A} = b(\rho)\le(\bar{a}(x^+,x^-) + d\,\ri)b(\rho)^{-1}\,,  \qquad b(\rho)= e^{-\rho L_0}\,.
\end{align}
 Here $\rho$ is the holographic radial direction, and $x^+,x^-$ are boundary coordinates.  The connections \eqref{eq:aba} are a solution of the Chern-Simons equations of motion \eqref{eom} as long as
\be\label{eoma}
d{a} + {a} \wedge {a}=0\,, \qquad \quad d{\bar a} + {\bar a} \wedge {\bar a}=0\,.
\ee
The advantage of the parametrization in \eqref{eq:aba} is that we can isolate ($a$, $\bar a$), which act as the boundary values of the connections, and interpret the radial direction as emergent from a gauge transformation. The set of boundary conditions proposed in \cite{Banados:1998gg} leads to
\begin{align}
a_{x^+}&=\left(L_+-{2\pi {\cal L}(x^+)\over k}\,L_-\right),   &  a_{x^-}&=0\,, \nonumber\\
 \bar a_{x^+} &=0 \,, & \bar{a}_{x^-}&=-\left(L_--{2\pi \bar {\cal L}(x^-)\over k}\,L_+\right)\,.
\label{sl2conn}
\end{align}
where  ${\cal L}(x^+)$, and $\bar {\cal L}(x^-)$ are generic functions will always fulfil the equations of motion \eqref{eoma}. The connections described by  \eqref{eq:aba} with \eqref{sl2conn}  are known as {\it Ba\~{n}ados solutions}. These connections parametrize the space of all solutions that are asymptotically AdS$_3$ with a trivial flat boundary .

%\footnote{The BC is that $a_{-}=0$, and $e=1$. The $\omega_i =0$ is a gauge freedom we have because: 1) Campoleoni says it, even I do not understand. 2) If I put  $e=1$ in \eqref{constraints3}, it gives me $R=0$, even if I put $\omega_i \neq 0$ (i.e. flat boudnary anyway)  }

\section{Generalized solutions in Chern-Simons theory }\label{newsol}

The Ba\~{n}ados solutions in \eqref{radial}-\eqref{sl2conn}  analysed in the previous section have boundary conditions  $A_{x^-}=\bar A_{x^+} = 0$. In this section, we depart from this choice and find solutions to $SL(2,\RR)\times SL(2,\RR)$ Chern-Simons gauge theories with generic boundary conditions. We also find a variational principle with Dirichlet boundary conditions for the gauge field. Here, all the analysis is purely done in the Chern-Simons formalism of 3d gravity. However, in Section \ref{compmet}, we will show that our solutions exactly reproduce the Fefferman-Graham expansion in metric formalism \eqref{FG}. 

%We would like to comment that general boundary conditions for AdS$_3$ Chern-Simons theories have been also studied in \cite{Grumiller:2016pqb}. However, their solutions do not have a regular Fefferman-Graham expansion \eqref{FG0} when translated to the metric formalism, and the metric contains off-diagonal terms such $g_{x^{\pm}z}$. They find that the asymptotic symmetry algebra associated to their gauge condition consists of two copies of affine $\mathfrak{sl}(2)_k$-algebras. Similar boundary conditions were analyzed in \cite{Cotler:2018zff}, but our treatment will be more systematic.

\subsection{The connections }\label{nontrisol}

Here we assume the same gauge choice of the radial direction as in \eqref{radial}, but we will go away from the solution \eqref{sl2conn} with trivial boundary. We consider from now on $\ell=1$. We propose the following  general boundary conditions for $a$ and $\bar a$ is:
\be\label{abaraa}
a_i= 2 e_i^+ L_+ -\, f_i^- L_- \,+\, \omega_i L_0\,,\quad\quad \bar a_i= f_i^+ L_+ -\,2 e_i^- L_- \,+\, \omega_i L_0\,,
\ee
where  the index $i=x^{+},\,x^-$. The only choice we made is to have equal $L_0$ components for both $a$ and $\bar a$
%, exploiting the residual gauge freedom. 
The equations of motion \eqref{eom}, for each component in the Lie algebra indices, impose the following constraints for the parameters in \eqref{abaraa}
 %
%\footnote{\textcolor{cyan}{We have used the equations \eqref{eoma} by component in the flat indices are 
%\be\label{flatac}
%da^m+{\varepsilon_{nl}^{\quad m}\over 2}\,  a^n\wedge a^l=0\,,
%\ee
%where $\varepsilon_{nl}^{\quad m}=\varepsilon_{nlk}\,\eta^{dk}$, and we defined $\varepsilon^{0+-} =1$. I think it is better not to say this because I have to define the epsilon with three indices, that I never use later. Moreover the formula is very ugly, with all this indices.} }
%:
%
\begin{align}\label{constraints}
d\omega - 4\,e^+ \wedge f^- &= 0 \,, & d\omega - 4 \,f^+ \wedge e^-&= 0  \,\\
de^+ - \omega \wedge e^+  &= 0\,, &  df^+ -\omega \wedge f^+ &= 0\, ,
\\
 dt^- + \omega \wedge f^-&= 0\,,  &  de^- + \omega \wedge e^-  &= 0\,.
 \label{constraints2}
\end{align}
The left column corresponds to the constraints deduced from $a$, and the left column from $\bar a$.  We can write the previous equations more compactly, adding and subtracting different combinations of them
%
%%
%\begin{align}
%e^a \wedge f_a &= 0\,,\\
%d\omega - 2\,\varepsilon_{ab} \,e^a \wedge f^b &= 0 \,,\label{constraints3}\\
%de^a - \varepsilon^a_{\,\,b}\, e^b \wedge \omega &= 0\,, \label{constraints4}\\
%df^a - \varepsilon^a_{\,\,b}\, f^b \wedge \omega &= 0  \label{constraints5}
%\end{align}
%%
\begin{align}\label{constraints3}
d\omega - 2\,\varepsilon_{ab} \,e^a \wedge f^b = 0 \,, \qquad\qquad\qquad\quad   e^a \wedge f_a = 0\\
\qquad de^a - \varepsilon^a_{\,\,b}\, e^b \wedge \omega = 0\,, \qquad\quad  df^a - \varepsilon^a_{\,\,b}\, f^b \wedge \omega = 0  \label{constraints4}
\end{align}
Here, we have defined a new 2-dimensional flat index that has values $a, b=\{+,-\}$.
The symbol $\varepsilon_{ab}$ is the Levi-Civita tensor defined in  \eqref{conv}.
\footnote{In this text, we use labels $ m, n, l ...$ for the $sl(2,\mathbb{R})$ indices that run over $\{0, +, -\} $. The letters $a, b, c ...$ refer to the two-dimensional flat index with possible values $\{+,-\}$. 
%Notice the latter is a subset of the former, i.e.  $m=(0,a)$.
}
The appearance of this 2-dimensional indexing is a consequence of the gauge choice \eqref{radial}, where the radial dependence is taken care of $L_0$. Its origin will be clear in Section \ref{sec:viel}.
It is interesting to use \eqref{eq:aba},  together with \eqref{ident}, to analyse the full solution :
%%
%\be\label{abara}
%A_i= 2 e_i^+ e^{\rho}L_+ -\, f_i^- e^{-\rho}L_- \,+\, \omega_i L_0\,,\quad\quad \bar A_i= f_i^+ e^{-\rho} L_+ -\,2 e_i^- e^{\rho} L_- \,+\, \omega_i L_0\,,
%\ee
%%
%
\be\label{abara}
A_i= {2 e_i^+ \over z}L_+  - z\, f_i^-  L_- \,+\, \omega_i L_0\,,\quad\quad \bar A_i= z f_i^+ L_+ -\,{ 2 e_i^-\over z} L_- \,+\, \omega_i L_0\,.
\ee
where we have used the change $z=e^{-\rho}$ in the radial coordinate, with boundary located at $z\rightarrow 0$. 
We would like to remark that the solution \eqref{abara} is independent of the parametrization \eqref{eq:aba}. Alternatively, one could solve the equations of motion \eqref{eom} for $A$ and $\bar A$, with the gauge choice \eqref{radial}, and the most generic solution would also lead to \eqref{abara}, and truncate at some powers of $z$. 
The connections \eqref{abara} have generalized boundary conditions respect to the Ba\~nados solutions \eqref{sl2conn}. In Section \ref{compmet}, we will show that they reproduce the Fefferman-Graham expansion \eqref{FG} in Chern-Simons formalism.

\subsection{Boundary terms}\label{bdCS}

  %In Section \ref{nontrisol}, we have found most generic solution with Fefferman-Graham gauge of AdS$_3$ in Chern-Simons formalism.  We need to take care of boundary terms to ensure a well defined variational principle  for our solutions, i.e. $\delta S =0$. \textcolor{red}{(We will propose our own variational principle? Motivation? comparison)} We start by variating the Chern-Simons action \eqref{actiona}, which gives  
%%
%\be\label{actionabd}
%\delta S_{CS}[{A}]= \frac{k}{4\pi}\int_{\cal M}  \Tr\left({ F} \wedge \delta{A}\right) -\frac{k}{4\pi}\int_{\cal \partial M} \Tr\left({ A} \wedge \delta{A}\right)~.
%\ee
%%, and evaluated at $\cal \partial M$, the boundary of $\cal M$ 
%The first term gives the CS equations of motion, and the second is a boundary term obtained from partial integration.  We add the barred sector, and expand the wedge products using the light-cone coordinates at the boundary $(x^+,\, x^-)$. 
%
%----%
In pure 3-dimensional gravity,  the boundary conditions are especially important because we do not have bulk degrees of freedom and, therefore, all the interesting information is located at the boundary.  In Section \ref{nontrisol}, we have fixed a radial gauge for our solution, and now we need to make sure we can tie it to a well defined variational principle at the boundary. This is crucial in the context of the AdS/CFT correspondence, since AdS action on-shell is related to the CFT partition function. We start by varying the CS action in \eqref{actiona}, which  on-shell is
\be\label{actionabd2}
\delta S_{CS}[{A}]- \delta S_{CS}[{\bar A}]=-\frac{k}{4\pi}\int_{\partial {\cal M}} \Tr\left(A \wedge \delta A-\bar  A \wedge \delta  \bar A\right)~. 
\ee
where $\partial {\cal M}$ is the boundary of the manifold ${\cal M}$. We consider light-cone coordinates  $(x^+,\,x^-)$ for the boundary as in the previous sections,  and we can expand the wedge products as: 
\be\label{actionabdwedge}
\delta S_{CS}[{A}]- \delta S_{CS}[{\bar A}]=-\frac{k}{4\pi}\int dx^+dx^- \Tr\left(A_{x^+}\delta A_{x^-} -A_{x^-}\delta A_{x^+}   -\bar A_{x^+}\delta \bar A_{x^-} +\bar A_{x^-}\delta \bar A_{x^+} \right)~. 
\ee
For the Ba\~nados connections in \eqref{sl2conn}, we have $A_{x^+}=0$, and $\bar A_{x^-}=0$. Therefore, the variation of the action \eqref{actionabdwedge} is zero if we also consider $\delta A_x^+$ and $\delta \bar A_{x^-}$ fixed at the boundary, and hence we would have a well-defined variational principle. However, this argument does not hold for \eqref{abara}, because we have all non-zero components of the connection at the boundary. We need to define a new variational principle for these background solutions. We begin by expressing the variation of the action \eqref{actionabd2} in terms of the parameters in \eqref{abara}
\be\label{actionabd3}
\delta S_{CS}[{A}]- \delta S_{CS}[{\bar A}]=-\frac{k}{2\pi}\int_{\partial {\cal M}} \varepsilon_{ab}\left(e^a \wedge \delta f^b-f^a \wedge \delta e^b\right)~,
\ee
where the parameter $\omega_i$ cancels out due to the negative sign between the barred and non-barred sector of the actions.
%\textcolor{blue}{($\omega$?)}.  
It also remarkable that there is no explicit dependence on the radial coordinate $z$, unlike in the metric formalism, where the action needs to be renormalized, due to divergences at the boundary (see Section \ref{rev:met}). This is due to the gauge choice \eqref{radial}, and the subsequent gauge reparametrization \eqref{eq:aba}. Note, however, that the connections \eqref{abara} still diverge at the boundary with
\be\label{abara2}
A_i = {2 e_i^+ \over z}\,L_+ +... \,,\qquad\qquad \bar A_i = -\,{ 2 e_i^-\over z}\, L_-+...\,.
\ee
where the dots stand for less dominant terms as $z\rightarrow 0$. We define a variational principle with Dirichlet boundary conditions for the connections, which will remain fixed at the boundary. This means that the variation of  parameter $e^a$ is null, and $f^a$ is free to vary.  With these considerations, the variation \eqref{actionabd3} is not zero, but we have the freedom to add a new boundary term to the action that accomplish it. We propose 
\bea\label{varf}
S=  S_{CS}[{A}]- S_{CS}[{\bar A}]+S_{\text{bd}}\,,
 \eea
where $S_{\text{bd}}$  is defined as: 
%\footnote{\textcolor{red}{Can I contract with this 2d levi-civita? if I am still in 3d, and I have not gave any meaning. Covariant like?} }
%
\bea\label{actionbdvar}
 S_{\text{bd}}=-\frac{k}{2\pi}\int_{\cal \partial M} \epsilon_{ab}\left({ A^a} \wedge {A^b}+ {\bar A^a} \wedge {\bar A^b}\right)~.
\eea 
%
%\bea\label{actionbdvar}
%\delta S_{\text{bd}}=-\frac{k}{4\pi}\int_{\cal \partial M} \epsilon_{ab}\left({ A^a} \wedge \delta{A^b}+ {\bar A^a} \wedge \delta{\bar A^b}\right)~.
%\eea 
% %
The term $ S_{\text{bd}}$ is designed such that the total action has zero variation when $e^a$ is held fixed, and, in fact, we see that
%
%\footnote{We see that the variation of this action expanded is
%\be\label{actionabd3}
%\delta S_{\text{bd}}=\frac{k}{2\pi}\int_{\partial {\cal M}} \varepsilon_{ab}\left(e^a \wedge \delta f^b+f^a \wedge \delta e^b\right)~. 
%\ee
%}
%
\be\label{actionabd4}
\delta S = \frac{k}{\pi}\int_{\partial {\cal M}} \varepsilon_{ab}\,f^a \wedge \delta e^b~. 
\ee
We have found then a variational principle for the connections \eqref{abara} with Dirichlet boundary conditions. This is necessary to interpret the connections holographically. The AdS/CFT dictionary tells us that we can identify \eqref{actionabd4} with the variation of the CFT the action, which schematically is $\int d^2x \,O(x) \,\delta J(x)$ \cite{Witten:1998qj}. Therefore, we see that  the vielbein $e^a$ acts as a source, whose variation vanishes at the boundary, and $f^a$ as the expectation value the dual operators. 

%We would like to point out that our variational principle is different from the one used for \eqref{sl2conn}. As explained in \eqref{actionabdwedge}, this is due to the fact that $a_{x^-}=\bar a_{x^+}=0$, for the Ba\~nados connections. In higher spin gravity, it is usual to consider connections with no null boundary conditions for $a_{x^-}$, $\bar a_{x^+}$, as it occurs for our case \eqref{abara}. However, our identification of sources and vevs also differs from the one that is normally used in the higher spin literature. See, for example, \cite{deBoer:2013gz}. In \cite{Miskovic:2006tm}, they also consider boundary conditions with no nul $a_{x^-}$, $\bar a_{x^+}$. In their case, they 

We would like to remark that our variational principle is different from the one used for the  Ba\~nados solutions \eqref{sl2conn}. As explained in \eqref{actionabdwedge}, this is due to the form of the connections \eqref{abara} which have $A_{x^-},\,\bar A_{x^+}\neq 0$ at the boundary. In \cite{Miskovic:2006tm}, they also analyse the variation of the Chern-Simons action \eqref{actionabd2} for connections with all components not null at the boundary. However, their definition of variational principal is different from ours, and they do not need to add any counterterm such our \eqref{actionbdvar}. 
%So, it means that they consider $\delta f$=0 at the boundary. They effectively do the variational principle for the boundary metric $h_{ij}$, not its expansion in g0, g2, and g4. See they consider only de CS original variation (44), which is equal (13), and null for $\delta K_i^j=0$, and $\delta e_i =0$. This only happens for the metric $h_{ij}$. For its expansion I need to consider extra terms
In higher spin gravity, it is as well usual to consider connections with no null boundary conditions for $a_{x^-}$, $\bar a_{x^+}$. Our identification of sources and vevs also differs from the one that is normally used in the higher spin literature. See, for example, \cite{deBoer:2013gz}.

\section{Solutions with non-trivial curved boundary}\label{compmet}

In this section, we show that the Chern-Simons connections \eqref{abara} represent geometries in the Fefferman-Graham expansion, which are the most generic solutions with boundaries that are asymptotically AdS$_3$. We first review these solutions in metric formalism in Section \ref{rev:met}. In Section \ref{sec:viel}, we identify the Chern-Simons parameters in \eqref{abara} as CFT quantities:  $e_i$, $\omega_i$, and $f_i$ are related to a 2-dimensional vielbein, spin connection, and stress tensor, respectively.  

We would like to comment that general boundary conditions for AdS$_3$ Chern-Simons theories have been also studied in \cite{Grumiller:2016pqb}. However, their solutions do not have a regular Fefferman-Graham expansion \eqref{FG0} when translated to the metric formalism, and the metric contains off-diagonal terms such as $g_{x^{\pm}z}$. They find that the asymptotic symmetry algebra associated to their gauge condition consists of two copies of affine $\mathfrak{sl}(2)_k$-algebras. Moreover, similar boundary conditions to \eqref{abaraa} have been discussed in  \cite{Cotler:2018zff}. They also relate the dominant term in their connections to a 2-dimensional vielbein and find a well-defined variational principle at the boundary. However, our identification of the parameters in the connections \eqref{abara} as CFT quantities is more systematic, which will be useful for future applications. For example, the relation we find between $f_i$ and the boundary stress tensor will be crucial in Section \ref{sec:DTdef} for the analysis of double-trace deformations. 

\subsection{Review: metric formulation}\label{rev:met}

%Conventions taken from skenderis, paper Nov 2000 (1.1)
Let us start by considering a set of coordinates where the metric is in {\it Fefferman-Graham gauge}:
\be\label{FG}
ds^2=\ell^2{dz^2\over z^2} +g_{ij}(x^k, z)dx^idx^j,   
\ee 
Solutions to Einstein's field equations with negative cosmological constant truncate in three dimensions \cite{Skenderis:1999nb}, which gives
\be\label{AdSassymp}
g_{ij}(x^k,z)={g_{ij}^{(0)}(x^k)  \over z^{2}}+g_{ij}^{(2)}(x^k)+ z^{2}\,g_{ij}^{(4)}(x^k)\,,
\ee
where $g_{ij}^{(0)}$ is interpreted as the metric of the boundary theory.  The elements of $g^{(4)}_{ij}$ are fully determined  in terms of $g_{ij}^{(2)}$, and $g_{ij}^{(0)}$ by:
% \be\label{g4g2}
% g^{(4)}_{\alpha\beta}=\frac{1}{4} g^{(2)}_{\alpha\gamma} g^{(0) \, \gamma\delta} g^{(2)}_{\delta\beta}\,,
% \ee
  \be\label{g4}
 g^{(4)}_{ij}=\frac{1}{4} g^{(2)}_{i k} g^{(0) \, k l} g^{(2)}_{l j}\,. 
 \ee
 Moreover, $g^{(2)}_{ij}$ fulfils the following conditions: 
 \be\label{g2}
g^{(2)\,\,i}_{\,\,\,\,\,\,\,\,i}= \text{ tr}\left((g^{(0)})^{-1}g^{(2)}\right)= {\ell^2 \over 2} R^{(0)}\,,\qquad\qquad  \nabla^{(0)\,i} g^{(2)}_{ij}=\nabla_{j}^{(0)} g^{(2)\,\,i}_{\,\,\,\,\,\,\,\,i}\,.
 \ee
The indices have been raised and lowered with the metric $g^{(0)}_{ij}$, and $R^{(0)}$ and  $\nabla_{j}^{(0)}$, are the Ricci scalar, and the covariant derivative in the metric $g^{(0)}_{ij}$.  In \cite{Balasubramanian:1999re}, they interpret these geometries holographically. For that, it is necessary to define a variational principle with Dirichlet boundary condition for the metric: when $g_{ij}^{(0)}$ is held fixed, and the subleading terms can vary at the boundary.  Initially, the metric  \eqref{AdSassymp} is not an extremum of the Einstein-Hilbert action, because its variation contains linear terms in $\delta g^{(2)}_{ij}$, and divergences at the boundary. To circumvent this problem, it is necessary to add an appropriate counterterm:
\begin{align}\label{GHterm}
 S_{\text{ct}}=-{1\over 8 \pi G_3}\int_{\partial \cal M}d^2x\sqrt{-g}\,,
 \end{align}
For a review on this analysis, see e.g. \cite{Kraus:2006wn}. After adding \eqref{GHterm} to the Einstein-Hilbert action, with a Gibbons-Hawking term, the on-shell variation of the total action is:
%In \cite{Balasubramanian:1999re}, they propose, the Gibbons-Hawking boundary term, and a counter-term \textcolor{blue}{(I do not need to show the following boundary terms for anything. I could skip them in principle, and just jump to  \eqref{stressmet}, while properly citing. Consider if they are necessary to undertsand the difference in the divergence. However, not adding them it is nicer for the flow to the stress tensor anomaly. )}:
%  \begin{align}\label{GHterm}
% S_{\text{GH}}&={1\over 8 \pi G_3}\int_{\partial \cal M}d^2x\sqrt{-g}\Tr\left(K\right)\,,\qquad
% S_{\text{ct}}=-{1\over 8 \pi G_3}\int_{\partial \cal M}d^2x\sqrt{-g}\,,
%  \end{align}
%where $K_{ij}={\partial_{\rho}g_{ij}/ 2}$ is the extrinsic curvature, and  $\partial \cal M$ is the boundary of the 3-dimensional manifold $\cal M$. The terms \eqref{GHterm} ensure that the action
%\be
%S= S_{\text{EH}} +S_{\text{GH}}+ S_{\text{ct}}\,,
%\ee
 \be\label{stressmet}
 \delta S={1\over 2}\int_{\partial \cal M}d^2x\sqrt{-g^{(0)}}T^{ij}\delta g_{ij}^{(0)}\,,
 \ee
%\footnote{I used the convention in Per notes which are for euclidean, but I assumed it is the same. Because in per-vijay is in lorentzian and seems the same.} 
where the boundary stress tensor $T_{ij}$ is defined as:
    \be\label{stressBY}
  T_{ij}={1\over 8 \pi G_3\ell}\left(g_{ij}^{(2)}- g^{(2)\,\,k}_{\,\,\,\,\,\,\,\,k}\, g_{ij}^{(0)}\right)\,,
  \ee
 It is interesting to see that the counterterm \eqref{GHterm} is divergent at the boundary when we consider the expansion \eqref{AdSassymp}, as opposed to what happened in the boundary terms in the Chern-Simons formalism studied in Section \ref{bdCS}. The quantity $T_{ij}$ is a symmetric tensor that obeys the properties of stress-energy tensor on the CFT  \cite{Balasubramanian:1999re}. For example, using \eqref{g2} we find
 \be\label{Weyl}
 \nabla^i  T_{ij}=0\,,\qquad \qquad T_i^i = -{\ell\over 16 \pi G_3} R^{(0)}\,.
 \ee
 The first equality shows that  $T_{ij}$  is conserved, and the second recovers the CFT Weyl anomaly,  $T_i^i=-{c\over 24 \pi} R$,  from which we can deduce the central charge $ c={3\ell \over 2 G_3}$. This illustrates that $g^{(0)}_{ij}$ is the metric of the CFT$_2$ living on the boundary of the the AdS$_3$.

  %Moreover, in \cite{Balasubramanian:1999re}, they derive that  transforms under diffeomorphisms as a tensor plus a Schwarzian derivative. 
 
% Let us assume that the boundary metric is $g^{(0)}=-dx^+dx^-$. In this case, Einstein's equations imply:
%  \be
%g_{+-}^{(2)}=g_{-+}^{(2)}=0\,,\qquad \partial_- g_{++}^{(2)}=\partial_+ g_{--}^{(2)}=0\,.
% \ee
%In this case, the stress tensor is traceless and conserved stress tensor at the boundary $g^{(0)}$, and has only two non-vanishing components:
%\be\label{3dstress}
%T_{++}={1\over 8 \pi G_3\ell}g_{++}^{(2)}(x^+)\,,\qquad T_{--}={1\over 16 \pi G_3\ell}g_{--}^{(2)}(x^-)\,.
%\ee
% Using the boundary stress energy tensor we can compute conserved charges in gravity \cite{Brown:1992br}. 
%
\subsection{Connections with asymptotically curved boundary}\label{sec:viel}

%\subsubsection*{Boundary vielbein and spin connection}

In this section, we show that the asymptotically AdS$_3$ metric solutions presented in Sec. \ref{rev:met} are equivalent to the Chern-Simons connections studied in Sec. \ref{nontrisol}. The first indication is that we have chosen the same gauge conditions: the choice \eqref{radial} is equivalent to the Fefferman-Graham gauge in \eqref{FG}.
\textcolor{blue}{($\omega$)}. 
One can easily see this using \eqref{metsl2} to relate the metric and the connections. Moreover, we will now show that the parameters $e_i$ and $\omega_i$ in \eqref{abara} represent the vielbein and the spin connection of the theory living on the curved boundary. We start seeing this, identifying $g^{(0)}$ in \eqref{metsl2} from the expansion \eqref{AdSassymp}:
\be\label{abrevmet}
g_{ij}^{(0)}= -2 (e^+_i e^-_j +e^-_i e^+_j  )=2 \,e_{i}\cdot e_{j} \,.
\ee
where in the second equality we have defined the scalar product as an abbreviation for the contraction with the metric $\eta_{ab}$, defined in \eqref{conv}. Equation \eqref{abrevmet} resembles the definition of a vielbein \eqref{vielbein1}, up to a normalizations of the overall flat metric. Since $g_{ij}^{(0)}$ is the boundary metric, we identify $e^a_i$ as a {\it boundary vielbein}. The boundary vielbein $e^a_i$ should not be confused with the 3-dimensional bulk vielbein $E_{\mu}^m$ used in Section \ref{sec:CS}, see Appendix \ref{sec:conv}. We can see that $e_i$ controls the dominant term of  $E_i$, as defined from \eqref{vbspin}:
\be\label{bdviel}
 E_{i}= {A_{i} -\bar A_{i}\over 2}  = \left(e^+_{i} \, L_+\,+\,e^-_{i} \, L_-\right) \, {1\over z } + \dots
\ee
where the dots mean we ignored lower orders in $z$. Moreover, the right constraint in \eqref{constraints4}  is equivalent to the torsion equation \eqref{torsion} in 2-dimensions, where the spin connection \eqref{spinc} has only one independent component. Therefore, if $e_i$ is the boundary vielbein, the equation in the right of \eqref{constraints4} defines the parameter $\omega_i$ as the {\it boundary spin connection}. The Ricci scalar in  2 dimensions is \eqref{Ricci}, which shows that the solution \eqref{abara} describes theories with non-trivial curved boundaries, in contrast with Ba\~nados connections \eqref{sl2conn}.

We have seen that the connections \eqref{abara} and the metric \eqref{AdSassymp} are both in Fefferman-Graham gauge, and allow for boundary theories with non-trivial metric. However, to ultimately check that the two solutions are the same but in different formalisms, we now prove that the the equations of motion \eqref{constraints3}-\eqref{constraints4} are equivalent to the constraints \eqref{g4}-\eqref{g2}.  Using \eqref{metsl2}, we can find $g^{(2)}$, and $g^{(4)}$ in terms of the parameters of the connections:
\be\label{abrevmet2}
 g_{ij}^{(2)}=-2 e_{i}\cdot f_{j}=-2 f_{i}\cdot e_{j}\,,\qquad g_{ij}^{(4)}={1\over 2} f_{i}\cdot f_{j}\,,
\ee
where we have used the right equation in \eqref{constraints3} to write the expression for $g_{ij}^{(2)}$.  We start reproducing the constraint \eqref{g4} with%
\begin{align}
\frac{1}{4} g^{(2)}_{i k} g^{(0) \, k l} g^{(2)}_{l j}\,=\,2\,  \varepsilon^{kk'}\varepsilon^{ll'} (f_{i}\cdot e_{k} ) ( e_{k'}\cdot e_{l'}) (e_{l}\cdot f_{j})\,=\,  {1\over 2} f_{i}\cdot f_{j} \label{g4pr}  =g^{(4)}_{ij}\,,
\end{align}
 where $\varepsilon^{ij}$ is the levi-civita tensor with curved indices defined in \eqref{levicg}. In the first equality, we used the expression in \eqref{invg0} for the inverse of $g^{(0)}_{ij}$. For the second equality, we used the relation \eqref{contree}. The left constraint in \eqref{g2} is recovered by
 \be\label{proofg2}
 \text{ tr}\left((g^{(0)})^{-1}g^{(2)}\right) = \,-4\,  \varepsilon^{ii'}\varepsilon^{jj'}( e_{i'}\cdot e_{j'}) (e_{i}\cdot f_{j})\,=- \varepsilon^{ij} \partial_{i}\,\omega_{j}  = {R^{(0)}\over 2}\,.
 \ee 
 For the second equality we used \eqref{contree}, and the equation in the left of \eqref{constraints3}, and for the last equality, the definition of the Ricci scalar in \eqref{Ricci} in two dimensions. We recover the first constraint in  \eqref{g2} with $\ell=1$. %\textcolor{blue}{(Say explicitly where we have set $\ell=1$, in the CS) } 
 The right constraint in \eqref{constraints3} can be written in terms of the covariant derivative \eqref{covariant} as:
 \be
 \nabla_{[i}f^a_{j]}=0\,.
 \ee
The previous equation, together with \eqref{tetradpost}, are enough to reproduce the differential condition over  $g^{(2)}$ in \eqref{g2}.

%\textcolor{blue}{To finally write/clean this section, I need the structure of this section). 1) To prove that both theories are equivalent, we need to compare the boundary terms in the action. (Bd conditions are as important as eom. But it might not be the case that boundary terms are as important as solutions. Because solutions does not exist without bd terms.) 2) Or we can just say we want to compare both variational principles... }

In  Section \ref{bdCS}, we found a variational principle for the solutions \eqref{abara} that keeps the boundary value of the CS connections fixed. We will now show that this variational principle is equivalent to the one in the metric formulation reviewed in Section \ref{rev:met}. To start the comparison, we translate the term \eqref{stressmet} to the vielbein formalism using \eqref{abrevmet}: 
\be\label{seT}
\delta S= 4\int_{\cal \partial M}\,d^2x e \, T_a^{i} \, \delta  e^a_{i}\,,
\ee
where we used the relation between determinants below \eqref{levicg}. Comparing \eqref{actionabd4} with \eqref{seT}, we can identify the components of the stress-energy tensor. Expanding the wegde product, its compact expression turns out to be 
\be\label{stressTmine}
T_{a}^{i}=\frac{k}{\pi}\varepsilon_{ab} \varepsilon^{ij}f^b_{j}\,.
\ee
 It is interesting to compute the trace of the stress tensor:
\be\label{RicciCS}
T^i_i=   T_a^i\,e^{a}_{i} =\frac{ k}{2\pi} \varepsilon^{ij} \partial_{i}\,\omega_{j}  = -{1\over 16 \pi G_3} R^{(0)}\
\ee
where we have used the left constraint in \eqref{constraints4}, and identification of Chern-Simons level \eqref{eq:level}, and with $\ell=1$.
% \textcolor{blue}{$\ell=1$}
 This is the CFT Weyl anomaly we have recovered in the metric formulation in \eqref{Weyl}. Moreover,  we can show that \eqref{stressBY} is equal to \eqref{stressTmine} by changing to curved indices appropriately  contracting with the vielbein. Therefore, the definitions of the variational principle coincide in the metric and Chern-Simons solutions. To finish, we would like to comment that, as expected, the Ba\~nados connections \eqref{sl2conn} represent solutions with flat metric at the boundary, and have null Weyl anomaly.

\section{$T\bar T$-deformation on the boundary of Chern-Simons AdS$_3$ gravity}\label{sec:DTdef}

In this section, we use the general AdS$_3$ solutions \eqref{abara} in Fefferman-Graham gauge, and the variational principle defined in \eqref{actionabd4} to study double-trace operators in the dual theories of $SL(2,\RR)\times SL(2,\RR)$ Chern-Simons gauge theory. We will analyze a specific deformation in the boundary of the Chern-Simons theory that will turn out to be $T\bar T$-deformations in dual CFT.

\subsection{Mixed boundary conditions in Chern-Simons formalism}\label{doubleCS}

In Section \ref{bdCS}, we found a variational principle with Dirichlet boundary conditions for the CS connections \eqref{abara}. This allows us to interpret the Chern-Simons boundary action \eqref{actionabd4} in the CFT, with the quantities $e^a$ and  $f^a$ acting as sources and expectation values of the dual operator, respectively.  In this section, we will use this knowledge to study the addition of double-trace combinations of $f^a$ to the boundary action. 
We focus on double-trace deformations of the type:
\be\label{TTCS}
S_{f^+f^-} ={1\over 2}\left( \frac{k}{\pi}\right)^2\int_{\partial{\cal M}}  f^- \wedge f^+\,,
\ee
So far, we have not motivated the choice of this double-trace operator, but it will become clear in Section \ref{ffastt}. It is well known \cite{Klebanov:1999tb,Witten:2001ua} that performing double-trace deformations in a CFT with large $N$ is equivalent to a change of  the boundary conditions of the dual bulk quantities. From the variational principle, one can find  modified relations between the asymptotic value of the bulk fields, and the sources and the expectation values \cite{Gubser:2002vv,Papadimitriou:2007sj}. For a more recent review of the procedure, see \cite{Bzowski:2018pcy}. We start by writing the variation of the action in the deformed theory as the variation of a new on-shell action, where the sources and operators depend on the deformation. In our case, this reads 
  \be\label{defvp}
 \delta S(0) +\lambda \, \delta S_{f^+f^-}(0)=  \delta S(\lambda)= \frac{k}{\pi}\int_{\partial {\cal M}} \varepsilon_{ab}\,f^a (\lambda)\wedge \delta e^b(\lambda)\,,
 \ee
where $ \delta S$ is the action of the original theory \eqref{actionabd4}, and $\lambda$ is the deformation parameter. The label inside the parenthesis indicates if the quantities belong to the original or the deformed theory. We need to solve \eqref{defvp} for $f^a (\lambda)$ and $e^b(\lambda)$ to find how the vevs and sources change in term of the deformation parameter. This is equivalent to solving the following equation:
%\textcolor{red}{(explain better the flow equation)}
\be\label{flowor}
{\partial \over \partial \lambda} \delta S(\lambda)  = - \delta S_{f^+f^-} (\lambda)
\,,
\ee
which gives
 %
%\begin{align}\label{prefloweq}
%\varepsilon_{ab}\left({\partial e^{a}\over\partial{\lambda}}\wedge  \delta f^{b}  +  e^{a}\wedge  \delta\left({\partial f^{b}\over\partial{\lambda}}\right)\right) \quad={k\over \pi} \left( \delta f^- \wedge  f ^+\, +\, f^- \wedge \delta  f ^+\right) \,, 
%\end{align}
%%
\begin{align}\nonumber
\varepsilon_{ab}{\partial e^{a}\over\partial{\lambda}}\wedge  \delta f^{b}  + \varepsilon_{ab}\, e^{a}\wedge  \delta\left({\partial f^{b}\over\partial{\lambda}}\right) &\quad=\quad-{k\over 2\pi} \left( \delta f^- \wedge  f ^+\, +\, f^- \wedge \delta  f ^+\right) \\
 &\quad=\quad\,\,\,{k\over 2\pi} \varepsilon_{ab} f^a \wedge \delta f^b \,,\label{prefloweq}
\end{align}
%
%$e^a\equiv e^a(\lambda)$, and$f^a\equiv f^a(\lambda)$
where $e^a$, and $f^a$ depend on the deformation parameter $\lambda$, but we leave the dependence implicit in this equation for simplicity in the notation. The second line in \eqref{prefloweq} allows us to see that \eqref{flowor} is solved by: 
\be\label{floweq}
{\partial e_i^{a}(\lambda)\over\partial{\lambda}}= {k\over 2\pi}  f_i^a (\lambda)\,,\qquad \qquad {\partial f_i^{a}(\lambda)\over\partial{\lambda}}= 0\,.
\ee
These are the equations for the flow of the CS parameters under a $T\bar T$-deformation. The flow of the spin-connection $\omega$ can be found from the constraints in \eqref{constraints4} using the flow of the other CS parameters. As we found, the left equation in \eqref{constraints4} is the torsionless condition \eqref{torsion} in 2 dimensions in vielbein formalism, whose solution we know for the spin connection \cite{Strobl:1999wv}:
%%
%\be\label{spinconne}
%\omega_i=   e \,e_{ia} \varepsilon^{jk}\partial_{j}e_k^a\,.
%\ee
%%
%
\be\label{spinconne}
\omega_i(\lambda)=e_{ia}(\lambda)\,(* d e^a(\lambda))\,.
\ee
where the symbol $*d e^a$ represents the Hodge dual of the 2-form $d e^a$. 

%blah blah. solve. 

Now, we consider the following  initial conditions at the beginning of the flow:
\be
e^a_i(0)=  e^a_i\,,   \qquad \quad f^a_i(0)=  f^a_i\,,   \qquad \quad \omega_i(0)=  0\,,
\ee
which means we have choose a flat theory at the boundary, because the Ricci scalar \eqref{RicciCS} is zero. It is easy to solve the equations \eqref{floweq}, and with this boundary conditions: 
\be\label{flowsol}
 e^a_i(\lambda)= e^a_i+\lambda \frac{k}{2\pi}f^a_i\,,  \qquad\qquad  f^a_i(\lambda)=  f^a_i\,.
\ee
We see that the parameter $f^a(\lambda)$ is constant under the flow, and  $e^a(\lambda)$ has a linear behaviour in $\lambda$. As expected, after the addition of the deformation, the boundary conditions mix the sources and expectations of the expectation values. We can deduce the flow of spin-connection using \eqref{spinconne} with \eqref{flowsol}:
\be\label{spinflow}
\omega_i(\lambda)=\left(2e_{ai}+\lambda\frac{k}{2\pi}f_{ai}\right)\left(* d e^a+\lambda \frac{k}{2\pi}(* d f^a)\right) =0\,,
\ee
where in the last equality we have used that $d e^a=d f^a=0$, since the initial parameters $e^a$ and $f^a$ follow the constraints \eqref{constraints4} for $\omega_i=0$. We see that if we start with a theory with a flat boundary theory, it will remain flat after  evolution under a $f^+f^-$-deformation. 
%Before moving on, we would like to comment that  \eqref{abara}  flow as:
%
%\be\label{abara2}
%A\,|_{z\rightarrow 0}= { \left(e^+  -\lambda \frac{k}{\pi}f^+ \right)}\,{2 L_+ \over z} \,,\qquad\qquad \bar A\,|_{z\rightarrow 0} = -\,{  \left(e^- -\lambda \frac{k}{\pi}f^-\right)}\, {2 L_- \over z}\,.
%\ee
%%We cannot find a way to naturally rewrite \eqref{abara2} as an induced connections at certain cutoff. 

\subsection{ $f^+f^-$ as a $T\bar T$-deformation}\label{ffastt}

In Section \ref{doubleCS}, we analysed the flow of under the double trace deformation \eqref{TTCS}. Now, we will see that $f^+f^-$ is equivalent to the usual $T\bar T$ operator in CFT. Remember the definition \cite{Zamolodchikov:2004ce}:
% Let us start by reminding that the  $T\bar T$ operator in the CFT is defined as:
%
\be
T\bar T\equiv T^{ij} T_{ij}-T^2=-2 \varepsilon^{ab}\varepsilon_{ij}  T _a^{i} T _b^{j}\,,
%=  -{1\over 2 }\varepsilon_{ab}\varepsilon^{ij}  T ^a_{i} T ^b_{j}
\ee
where in the second equality we have conveniently written it in terms of the flat indices using the vielbein. We consider the action of this double-trace operator
\be\label{TTviel}
S_{T\bar T}=-\int_{\partial{\cal M}} d^2 x\, e\,  \varepsilon^{ab}\varepsilon_{ij}  T _a^{i} T _b^{j}\,,
\ee
and using the holographic expression for the stress tensor in \eqref{stressTmine} to write the this action in terms of the Chern-Simons parameters we obtain
\be\label{TTCS2}
S_{T\bar T}= -\left( \frac{k}{\pi}\right)^2\int_{\partial{\cal M}} d^2 x\,e\,  \varepsilon_{ab}\varepsilon^{ij}  f ^a_{i} f ^b_{j} ={1\over 2}\left( \frac{k}{\pi}\right)^2\int_{\partial{\cal M}}  f^- \wedge f^+\,,
\ee
Therefore the $T\bar T$-deformation  is equivalent to the is equivalent to the $f^+f^-$-deformations \eqref{TTCS}.
\\
\\
It is instructive to compare with the metric formalism. In  \cite{Guica:2019nzm}, they use the variational principle method to analyse the flow of operators and sources after a $T\bar T$-deformation in a CFT.  They then interpret this results holographically, and find that the metric field has mixed boundary conditions after the deformation.  Here, we will see that the mixed boundary conditions found in \cite{Guica:2019nzm}, are equivalent to \eqref{flowsol}. For that we use \eqref{abrevmet}, which we write as
\be\label{metflow}
g_{ij}^{(0)}(\lambda)=2 \,e_{i}(\lambda)\cdot e_{j} (\lambda)= 2 \,e_{i}\cdot e_{j} +2\frac{k\lambda}{\pi}f_i\cdot e_{j}+ {1\over2}\left(\frac{k\lambda}{\pi}\right)^2 f_i\cdot f_{j}\,.
\ee
Using the expressions \eqref{abrevmet} and \eqref{abrevmet2}, we can rewrite in terms of metric quantities in the original theory:
\be\label{metflow2}
g_{ij}^{(0)}(\lambda)=  \,g_{ij}^{(0)} -\frac{ k\lambda}{\pi}g_{ij}^{(2)}+ \left(\frac{k\lambda}{\pi}\right)^2g_{ij}^{(4)}\,.
\ee
Comparing with \eqref{AdSassymp}, we see that \eqref{metflow} can be interpreted as an induced metric at a surface of constant $z=z_c$, where $z=z_c$ is defined as
\be\label{cutoff}
z_c^2=- \frac{ k\lambda}{\pi}= -{\lambda \over 4 \pi G_3}
\ee
where we have used \eqref{eq:level} with $\ell=1$. The interpretation as induced metric at $z=z_c$ is only valid for when $\mu < 0$. With this, we recovered exactly the result found in \cite{Guica:2019nzm}. 
%To get to this result, they ... In our case, it is easier. We have found solutions, which look much more simpler than in metric formalism. 
Before moving on,  %from the point of view of the connections, this mixed boundary condition that appear in TTbar are. 
we observe that the boundary values of the connections \eqref{abara}  flow as:
\be\label{abara2}
A\,|_{z\rightarrow 0}= { \left(e^+  +\lambda \frac{k}{2\pi}f^+ \right)}\,{2 L_+ \over z} \,,\qquad\qquad \bar A\,|_{z\rightarrow 0} = -\,{  \left(e^- +\lambda \frac{k}{2\pi}f^-\right)}\, {2 L_- \over z}\,.
\ee
We cannot find a way to naturally rewrite \eqref{abara2} as an induced connections at a certain cutoff. It is surprising there is not a similar argument to \eqref{metflow2} to deduce the radial cutoff from the Chern-Simons boundary connections.

%In Chern-Simons formalism we can interpret the radial direction as  emergent from a gauge transformation due to the parametrization \eqref{eq:aba}. We expected this parametrization to be useful when studying conformal field theories with $T\bar T$-deformations, since they can be viewed as AdS$_3$ geometries at a finite radial cutoff. However, we see this is not the case if we observe how the boundary values of the connections \eqref{abara}  flow under a $T\bar T$-deformation:
%%
%\be\label{abara2}
%A\,|_{z\rightarrow 0}= { \left(e^+  +\lambda \frac{k}{2\pi}f^+ \right)}\,{2 L_+ \over z} \,,\qquad\qquad \bar A\,|_{z\rightarrow 0} = -\,{  \left(e^- +\lambda \frac{k}{2\pi}f^-\right)}\, {2 L_- \over z}\,.
%\ee
%We cannot find a way to naturally rewrite \eqref{abara2} as an induced connections at a certain cutoff. It is surprising there is not a similar argument to \eqref{metflow2} to deduce the radial cutoff from the Chern-Simons boundary connections. 

\subsection{$T\bar T$-deformation in terms of $A$, and $\bar A$}

%Even though we can consider any term which is invariant to the boundary symmetries, 
We would like to write \eqref{TTCS} in terms of the full connections \eqref{abara}.  In this way, we can analyse if the $T\bar T$-deformation has a natural interpretation in the CS formulation of 3d gravity. To recover \eqref{TTCS}, we need combination of $f^+$, and $f^-$, and we should consider a term that combines $A$, and $\bar A$. Even thought  this is a non-standard choice in the Chern-Simons formulation of 3d gravity, there is nothing that holds us from doing so. 
\footnote{See \cite{Balasubramanian:2002zh}, for an example of boundary condition that mixes both sectors. }
Let me start by considering the simplest combination of $A$, and $\bar A$:
\begin{align}\label{TTinCSz}
%\frac{k}{\pi}
\frac{k}{4\pi}\int_{\partial {\cal M}} \Tr\left(A \wedge \bar A\right)  = &-\frac{k}{\pi} {1\over z^2} \int_{\partial {\cal M}} e^+ \wedge e^-\, - \frac{k}{4\pi}{z}^2\int_{\partial {\cal M}} f^- \wedge f^+\,. 
%\nonumber \\
%\\
%= &{4\over z^2} \int_{\partial {\cal M}}d^2x\, e\, - { 2 z^2} \int_{\partial {\cal M}} d^2 x\,\det( f)\,.\nonumber
\end{align}
%
%where in the second equality we have expanded the wedge product by components, and considered the definition of determinant \eqref{dete}. 
The first term in \eqref{TTinCSz} is divergent at the boundary. It is interesting to note that the integrand in the first term is the determinant of the vielbein, i.e.  $e^+ \wedge e^-=2 e \,d^2x$. Therefore, it is equivalent to the divergent part of the counterterm \eqref{GHterm} added in the metric formalism, upon the expansion \eqref{AdSassymp}. Analogously to the metric formalism, we get cancel the divergence by adding an appropriate counterterm:
\be\label{ttb}
\frac{k}{4\pi}\int_{\partial {\cal M}} \Tr\left(A \wedge \bar A\right)-\frac{k}{4\pi}\int_{\partial {\cal M}} \varepsilon_{ab} \,A^a \wedge \bar A^b  = - \frac{k}{2\pi}{z}^2 \int_{\partial {\cal M}} f^- \wedge f^+\,.
\ee
This term is similar to \eqref{TTCS}, but it depends in the radial coordinate $z$. To have a proper interpretation of \eqref{ttb} as a $T\bar T$-deformation in the boundary, this should not be the case. We could use the identification  
 $z=z_c$ with \eqref{cutoff} to avoid this problem. Again, the symbol $\lambda$ in $z_c$, would act as deformation. However, as explained in \eqref{abara2}, the identification of this boundary cutoff is not natural from the point of view of the CS connections. 

Another candidate to reproduce the $T\bar T$ deformation in terms of the connections  $A$, and $\bar A$:
\begin{align}
S_{T\bar T}&=\int_{\partial{\cal M}} e\, d^2x\, \epsilon^{ik} \epsilon^{jl}\,\tr(A_iA_j)\tr(\bar A_k\bar A_l)   
\nonumber\\
&=  \qquad  \int_{\partial{\cal M}} \,  d^2x\, {1\over e} \eta_{ab} \eta_{cd}( A^a\wedge \bar A^c  )  ( A^b\wedge \bar A^d  )\,.
\end{align}
This term is designed to exactly reproduce \eqref{TTCS}, up to an overall normalization constant, for \eqref{abara} with $\omega_i=0$.  The advantage of this term is that we do not have explicit $z$-dependence, so it is natural to consider it as a pure boundary deformation. However, this advantage brings a more complicated combination of $A$, and $\bar A$, as a downside.  In particular, we need  an explicit dependence in the determinant of the vielbein.  This is similar to the what happens the Chern-Simon $U(1)$ gauge field in \cite{Kraus:2006nb}, where we need to couple to the boundary metric to find a well variation principle. However, it is strange from the point of view of Chern-Simons as a gravitational theory, since all the information about the metric should be encoded in the connections. 

We have proposed two terms that could perform a $T\bar T$-deformation in terms of the connections $A$, and $\bar A$. However, it is not clear how natural they are from the point of view of the Chern-Simons variables. 

%It is not clear which one is better to reproduce the TT with $A$, and $\bar A$. We proposed two terms, with some downsides, and positive sides. Further analysis is done in the conclusion

%However, the disadvantage of combining barred and unbarred sectors is that the boundary terms are not independent of $z$, as happened for \eqref{actionabd2} due to the the radial choice \eqref{eq:aba}.
%\begin{align}
%S_{T\bar T}&=\int_{\partial{\cal M}}{1\over e}\left(  \tr \left(A_{x^+}^2\right)\tr \left(\bar A_{x^-}^2 \right) +   \tr \left(A_{x^-}^2\right)\tr \left(\bar A_{x^+}^2 \right)-2\tr \left(A_{x^+}A_{x^-}\right)\tr \left(\bar A_{x^+}  \bar A_{x^-}\right)  \right) \\
%&=  \int_{\partial{\cal M}}(T^+_{x^+}T^-_{x^-}-T^+_{x^-}T^-_{x^+})\,. 
%\end{align}
%%

\section{Conclusions}\label{sec:con}

In this article, we studied $\slt \times \slt$ Chern-Simons  gauge theories in AdS$_3$, and we presented two main results. The first one is the solution of the Chern-Simons equations of motion with generalized boundary conditions  \eqref{abara}. We have shown that these connections are equivalent to the Fefferman-Graham expansion \eqref{AdSassymp} in metric formalism. Moreover, we proved that we can interpret the different components of the gauge field as CFT quantities:  $e_i$, $\omega_i$, and $f_i$ are related to a 2-dimensional vielbein, spin connection, and stress tensor, respectively.  We have done this by proposing a variational principle with Dirichlet boundary conditions for the gauge connections.  This allowed us to study how the boundary of the Chern-Simons theory behaves under a specific type double-trace deformation  introduced in \eqref{TTCS}.  Using the variational principle method, we have found the mixed boundary after the deformation \eqref{flowsol} in the Chern-Simons formalism. This is our second result.  We showed that this deformation corresponds to a $T\bar T$-deformation from the point of view of the CFT. 

 Higher spin theories of gravity are very straight-forwardly generalized from the description of AdS$_3$ gravity as Chern-Simons gauge theory.  The generalization ammounts to extend the gauge group from $SL(2,\mathbb{R})$ to $SL(N,\mathbb{R})$. In \cite{Campoleoni:2010zq}, they find solutions to these theories that are considered asymptotically AdS$_3$ with flat trivial boundary. Assuming \eqref{eq:aba} these are
 \be\label{eq:a12}
 a_{x^+} = L_1 + \sum_{s=2}^{N} J_{(s)}(x^+) W^{(s)}_{-s+1} ~,\quad  \bar a_{{x^-}} =- L_{-1} +\sum_{s=2}^{N} \bar J_{(s)} (x^-) W^{(s)}_{s-1} ~,
 \ee
 where $ J_{(s)}(x^+)$, and $ \bar J_{(s)}(x^-)$ are any arbitrary function, and $a_{x^-} = \bar{a}_{x^+}=0$.
Here   $\{L_0,L_{\pm1}\}$ are the generators of the $sl(2,\mathbb{R})$ subalgebra in $sl(N,\mathbb{R})$, and $W^{(s)}_j$ are the spin-$s$ generators with $j=-(s-1),...(s-1)$. It would be interesting to generalise these solutions to have more generic boundary conditions of the type \eqref{abara}, as we proposed in this paper for the $sl(2,\mathbb{R})$ case. This, together with a well-defined variational principle, could allow us to formulate double-trace deformations involving spinning fields in Chern-Simons theory. We leave this idea as material for future work. 

%We would like to emphasize that the analysis carried out in this work is purely classical  But quantum aspects are not working well.  and that the quantum aspects 

\section*{Acknowledgements}

It is a pleasure to thank A. Castro and M. Guica for very useful discussions on the topic, and comments on the manuscript. The author was supported by the ERC Starting Grant 679278 Emergent-BH.

\appendix

\section{Vielbein formalism }\label{sec:dviel}

This appendix is a very short review, and collection of useful identities, of the vielbein formulation of general relativity. For some more context, see, for example \cite{Carroll:1997ar}. 
\\
\\
In the vielbein formalism, the choice of coordinates $x^{i}$ on a manifold is replaced by the election of a local basis.  This basis is formed by the $d$-dimensional vectors, $e^a = e^a_idx^{i}$, known as vielbeins. We can relate the metric of a curved manifold $g_{ij}$ to a flat (non-coordinate) metric $\eta_{ab}$ using the vielbein:
\be\label{vielbein1}
g_{ij}=\eta_{ab}e^a_{i}e^b_{j}\,.
\ee
%
%\textcolor{blue}{(put a 2 in front, to match the main text)}
The vielbein is used to change the indices of the tensors from curved to flat, or viceversa. For covariant derivatives of tensors with flat indices, the role of the Christofel symbol is replaced by the spin connection, characterized  by $\omega^{\,\,a}_{i\,\,b}$. For example, the necessary contractions in the case of a tensor with mixed indices $X^a_{j}$ are:
\be\label{covariant}
\nabla_{i}X^a_{j}=  \partial_{i}X^a_{j} +\omega^{\,\,a}_{i\,\,b}X^b_{j}-\Gamma^{\lambda}_{ij}X^a_{\lambda}\,. 
\ee
The following identity is required by construction of the vielbein formalism:
\be\label{tetradpost}
\nabla_{i}e^a_{j}=  0\,,
\ee
which is known as the tetrad postulade. Notice that \eqref{tetradpost} defines the spin connection in terms of the Christoffel symbols, and the vielbein. Moreover, the zero torsion condition, i.e., $\Gamma^{\lambda}_{ij}=\Gamma^{\lambda}_{ji}$ can be written in vielbein formalism as:
\be
de^a+\omega^a_{\,\,\,b}\wedge e^b=0\,. \label{torsion} 
\ee
The condition over  the flat metric $\nabla_{i}\eta_{ab}=0$ imposes the so-called metricity condition:
\be\label{metricity}
\omega_{i ab}=-\omega_{i ba}\,.
\ee
where $\omega_{i ab}=\eta_{ac}\,\omega^{\,\,c}_{i\,\,b}$.  It is also useful to write the curvature tensor with flat indixes as:
\be
R^a_{\,\,\,b} =d\omega^a_{\,\,\,b}+\omega^a_{\,\,\,c}\wedge \omega^c_{\,\,\,b} \label{curv}\,, 
\ee
where we defined   $R^a_{\,\,\,b}=R^a_{\,\,\,b\,ij}dx^i\wedge dx^j$.

\section{Conventions and useful identities}\label{sec:conv}

The review of vielbein formalism made in Appendix \ref{sec:dviel} is generic to any dimension. However, in this paper we have considered two different specific types of vielbein theories: a 3-dimensional one representing the bulk, and 2-dimensional one in the boundary. %\textcolor{blue}{(escriure aqui la diferencia de notacio?)} 
This appendix summarizes our conventions for both cases, and collects identities that are relevant for our manipulations. 

%\textcolor{blue}{(probably change order in the two subsections. It makes more sense to put 2d first, because it is a subsector of 3d.) }

\subsection{Vielbein formalism in 2 dimensions}\label{sec:2dviel}

 In the main text we described a 2-dimensional theory in vielbein formarlism, whose curved indices are represented by $\{i,j\}=x^+,x^-$, and flat indices $\{a,b\}=+,-$. Our conventions for the flat metric $ \eta_{ab}$ are
\be\label{conv}
  \eta_{+-}=\eta_{-+}=-1\,.
\ee  
and zero otherwise.  In two dimensions, the metricity condition \eqref{metricity} imposes that 
\be\label{spinc}
\omega_i^{ab}=\varepsilon ^{ab} \omega_i\,.
\ee
where $\omega_i$ is the only independent component of the 2-dimensional spin connection. The symbol $\varepsilon ^{ab}$  is the Levi-Civita defined as
\be\label{conv2}
\varepsilon_{+-}=-\varepsilon_{-+}=1\,,
\ee
One can raise and lower indices of the Levi-Civita tensor with the flat metric, as, for example: $\varepsilon^a_{\,\,b}=\eta^{ab'}\varepsilon_{b'b}$. It is useful also to define $\varepsilon^{ij}$ as Levi-Civita tensor with curved indices:
\be\label{levicg}
\varepsilon^{x^+x^-}= -\varepsilon^{x^-x^+}={1\over  \sqrt{-g}}={1\over 2 e }\,.
\ee
where $g=\det(g)$ and $e=\det(e)$ are the determinants of metric and the vielbein, which are related via  $ \det(g)=-4\det(e)^2$. The definition of determinant we used, for a generic tensor $X^a_i$:
%which are related via  $ \det(g)=-4\det(e)^2$.  
%The definition determinant we use is:
%%
%\begin{align}\label{dete}
%\det(e)= \,{1\over 2} \varepsilon_{ab}\tilde\varepsilon^{ij} e^a_i e^b_j \,=  e^+_{x^+}e^-_{x^-}-e^+_{x^-}e^+_{x^-}
%\end{align}
\begin{align}\label{dete}
\det(X)= \,e\, \varepsilon_{ab}\varepsilon^{ij} X^a_i X^b_j \,=  X^+_{x^+}X^-_{x^-}-X^+_{x^-}X^+_{x^-}\,.
\end{align}
We use the Levi-Civita symbols to write the inverse of the metric as
\be\label{invg0}
g^{ij} = \varepsilon^{ii'} \varepsilon^{jj'} g_{i'j'}\,,
\ee
Other useful relations using the Levi-Civita symbols are the following contractions of vielbeins: 
\be\label{contree}
\varepsilon^{ij} e^a_i e^b_j = -{\varepsilon^{ab}\over 2}\,,\qquad \qquad    e^i_a e^{j\,b}\varepsilon^a_{\,\,\,b}  = -{2\varepsilon^{ij}}\,.
\ee
We can use them, for example, to write find a compact expression for the Ricci scalar from the Riemann curvature tensor \eqref{curv}. In 2d, using \eqref{spinc}, we have%
\be
R^a_{\,\,\,b} =\epsilon^a_{\,\,\,b}d\omega \label{curv2}\,   \qquad\Leftrightarrow\qquad    R^a_{\,\,\,b\,i j}  = \epsilon^a_{\,\,\,b}\partial_{[i}\,\omega_{j]}\,,
\ee
We find the trace of $R^a_{\,\,\,b\,i j} $  by contracting its indices with the vielbein: 
\be\label{Ricci}
R= e^i_a e^{j\,b} R^a_{\,\,\,b\,i j}  =-2 \varepsilon^{ij} \partial_{i}\,\omega_{j}
\ee
where we have used the contraction of vielbein in the right of \eqref{contree}.

\subsection{Vielbein formalism in 3 dimensions}\label{app:sl2}

In 3-dimensions, we used the following definitions for the vielbein and spin connection:
\be\label{viel3d}
E^m=E^m_{\mu}dx^{\mu}\,, \qquad \qquad  \Omega^m= {1\over 2}\varepsilon^{mnl}\Omega_{\mu n l} dx^{\mu}\,,
\ee
%
%via $\omega^a=\epsilon^{abc} \omega_{bc}/2$, and $\epsilon_{abc}$ the Levi-Civita tensor. 
where  the capital the letters $E$ and, $\Omega$ to make a distinction with the 2-dimensional case described in Section \ref{sec:2dviel}. To avoid confusion, we used also different nomenclature for the  3-dimensional coordinate indices $\{\mu,\,\nu\}$, and the flat indices $\{m,n,l\}=0,\,\pm1$, which in this case are directly related to the $sl(2,\mathbb{R})$ algebra:
 \begin{align}\label{sl2alg}
     [L_0,L_{\pm}] = \mp L_{\pm}\,,\qquad [L_1,L_{-1}] = 2L_{0}\,,
\end{align}
Our conventions for the fundamental representation of  $sl(2,\mathbb{R})$ are
 \begin{equation}\label{sl2fund}
L_0 = \begin{pmatrix}
    1/2  & 0    \\
      0 &  -1/2
\end{pmatrix}\,, \quad
L_1 =  \begin{pmatrix}
    0  & 0    \\
      -1 &  0
\end{pmatrix}\,, \quad
L_{-1} =  \begin{pmatrix}
    0  & 1    \\
      0 & 0 
\end{pmatrix}\,.
\end{equation}
The Lie algebra metric reads
\be\label{sl2metric}
\Tr({L_0L_0})=\frac{1}{2}\,,\qquad \Tr({L_+L_-})=\Tr({L_-L_+})=-1~.
\ee
A useful identity between the generators of the algebra is: 
\be\label{ident}
e^{-L_0 \alpha} L_{m}  e^{L_0 \alpha}= L_{m}  e^{m \alpha}\,.
\ee

%A generic and simple expression, but very useful for simplifications will be:
%\be\label{antisymLC}
%\varepsilon^{ij}  a_{[i} b_{j]}  =  2  \varepsilon^{ij}  a_{i} b_{j} 
%\ee

%\be\label{epseps}
%\varepsilon^{ab}\varepsilon^{cd}= \delta^a_c  \delta^b_d- \delta^a_d  \delta^b_c
%\ee

\bibliographystyle{JHEP-2}
\bibliography{HigherSpin}

\providecommand{\href}[2]{#2}\begingroup\raggedright\begin{thebibliography}{10}

\bibitem{Achucarro:1987vz}
A.~Achucarro and P.~K. Townsend, {\it {A Chern-Simons Action for
  Three-Dimensional anti-De Sitter Supergravity Theories}},  {\em Phys. Lett.}
  {\bf B180} (1986) 89.
%%CITATION = PHLTA,B180,89;%%

\bibitem{Witten:1988hc}
E.~Witten, {\it {(2+1)-Dimensional Gravity as an Exactly Soluble System}},
  {\em Nucl.Phys.} {\bf B311} (1988) 46.
%%CITATION = NUPHA,B311,46;%%

\bibitem{Aragone:1983sz}
C.~Aragone and S.~Deser, {\it {HYPERSYMMETRY IN D = 3 OF COUPLED GRAVITY
  MASSLESS SPIN 5/2 SYSTEM}},  {\em Class. Quant. Grav.} {\bf 1} (1984) L9.
%%CITATION = CQGRD,1,L9;%%

\bibitem{Blencowe:1988gj}
M.~Blencowe, {\it {A Consistent Interacting Massless Higher Spin Field Theory
  In D = (2+1)}},  {\em Class.Quant.Grav.} {\bf 6} (1989) 443.
%%CITATION = CQGRD,6,443;%%

\bibitem{Bergshoeff:1989ns}
E.~Bergshoeff, M.~Blencowe and K.~Stelle, {\it {Area Preserving Diffeomorphisms
  and Higher Spin Algebra}},  {\em Commun.Math.Phys.} {\bf 128} (1990) 213.
%%CITATION = CMPHA,128,213;%%

\bibitem{Henneaux:2010xg}
M.~Henneaux and S.-J. Rey, {\it {Nonlinear $W_{infinity}$ as Asymptotic
  Symmetry of Three-Dimensional Higher Spin Anti-de Sitter Gravity}},  {\em
  JHEP} {\bf 1012} (2010) 007 [\href{http://arXiv.org/abs/1008.4579}{{\tt
  1008.4579}}].
%%CITATION = ARXIV:1008.4579;%%

\bibitem{Campoleoni:2010zq}
A.~Campoleoni, S.~Fredenhagen, S.~Pfenninger and S.~Theisen, {\it {Asymptotic
  symmetries of three-dimensional gravity coupled to higher-spin fields}},
  {\em JHEP} {\bf 1011} (2010) 007 [\href{http://arXiv.org/abs/1008.4744}{{\tt
  1008.4744}}].
%%CITATION = ARXIV:1008.4744;%%

\bibitem{Brown:1986nw}
J.~D. Brown and M.~Henneaux, {\it {Central Charges in the Canonical Realization
  of Asymptotic Symmetries: An Example from Three-Dimensional Gravity}},  {\em
  Commun.Math.Phys.} {\bf 104} (1986) 207--226.
%%CITATION = CMPHA,104,207;%%

\bibitem{Brown:1992br}
J.~D. Brown and J.~W. York, Jr., {\it {Quasilocal energy and conserved charges
  derived from the gravitational action}},  {\em Phys. Rev.} {\bf D47} (1993)
  1407--1419 [\href{http://arXiv.org/abs/gr-qc/9209012}{{\tt gr-qc/9209012}}].
%%CITATION = GR-QC/9209012;%%

\bibitem{Henningson:1998gx}
M.~Henningson and K.~Skenderis, {\it {The Holographic Weyl anomaly}},  {\em
  JHEP} {\bf 07} (1998) 023 [\href{http://arXiv.org/abs/hep-th/9806087}{{\tt
  hep-th/9806087}}].
%%CITATION = HEP-TH/9806087;%%

\bibitem{Balasubramanian:1999re}
V.~Balasubramanian and P.~Kraus, {\it A stress tensor for anti-de {S}itter
  gravity},  {\em Commun. Math. Phys.} {\bf 208} (1999) 413--428
  [\href{http://arXiv.org/abs/hep-th/9902121}{{\tt hep-th/9902121}}].
%%CITATION = HEP-TH 9902121;%%

\bibitem{Skenderis:1999nb}
K.~Skenderis and S.~N. Solodukhin, {\it {Quantum effective action from the AdS
  / CFT correspondence}},  {\em Phys. Lett.} {\bf B472} (2000) 316--322
  [\href{http://arXiv.org/abs/hep-th/9910023}{{\tt hep-th/9910023}}].
%%CITATION = HEP-TH/9910023;%%

\bibitem{Klebanov:1999tb}
I.~R. Klebanov and E.~Witten, {\it {AdS / CFT correspondence and symmetry
  breaking}},  {\em Nucl. Phys.} {\bf B556} (1999) 89--114
  [\href{http://arXiv.org/abs/hep-th/9905104}{{\tt hep-th/9905104}}].
%%CITATION = HEP-TH/9905104;%%

\bibitem{Witten:2001ua}
E.~Witten, {\it {Multitrace operators, boundary conditions, and AdS / CFT
  correspondence}},  \href{http://arXiv.org/abs/hep-th/0112258}{{\tt
  hep-th/0112258}}.
%%CITATION = HEP-TH/0112258;%%

\bibitem{Grumiller:2016pqb}
D.~Grumiller and M.~Riegler, {\it {Most general AdS$_{3}$ boundary
  conditions}},  {\em JHEP} {\bf 10} (2016) 023
  [\href{http://arXiv.org/abs/1608.01308}{{\tt 1608.01308}}].
%%CITATION = ARXIV:1608.01308;%%

\bibitem{Cotler:2018zff}
J.~Cotler and K.~Jensen, {\it {A theory of reparameterizations for AdS$_3$
  gravity}},  {\em JHEP} {\bf 02} (2019) 079
  [\href{http://arXiv.org/abs/1808.03263}{{\tt 1808.03263}}].
%%CITATION = ARXIV:1808.03263;%%

\bibitem{McGough:2016lol}
L.~McGough, M.~Mezei and H.~Verlinde, {\it {Moving the CFT into the bulk with $
  T\overline{T} $}},  {\em JHEP} {\bf 04} (2018) 010
  [\href{http://arXiv.org/abs/1611.03470}{{\tt 1611.03470}}].

\bibitem{Kraus:2018xrn}
P.~Kraus, J.~Liu and D.~Marolf, {\it {Cutoff AdS$_{3}$ versus the $
  T\overline{T} $ deformation}},  {\em JHEP} {\bf 07} (2018) 027
  [\href{http://arXiv.org/abs/1801.02714}{{\tt 1801.02714}}].

\bibitem{Taylor:2018xcy}
M.~Taylor, {\it {TT deformations in general dimensions}},
  \href{http://arXiv.org/abs/1805.10287}{{\tt 1805.10287}}.

\bibitem{Hartman:2018tkw}
T.~Hartman, J.~Kruthoff, E.~Shaghoulian and A.~Tajdini, {\it {Holography at
  finite cutoff with a $T^2$ deformation}},  {\em JHEP} {\bf 03} (2019) 004
  [\href{http://arXiv.org/abs/1807.11401}{{\tt 1807.11401}}].

\bibitem{Caputa:2019pam}
P.~Caputa, S.~Datta and V.~Shyam, {\it {Sphere partition functions
  \textbackslash{}\& cut-off AdS}},  {\em JHEP} {\bf 05} (2019) 112
  [\href{http://arXiv.org/abs/1902.10893}{{\tt 1902.10893}}].

\bibitem{Chen:2018eqk}
B.~Chen, L.~Chen and P.-X. Hao, {\it {Entanglement entropy in
  $T\overline{T}$-deformed CFT}},  {\em Phys. Rev. D} {\bf 98} (2018), no.~8
  086025 [\href{http://arXiv.org/abs/1807.08293}{{\tt 1807.08293}}].

\bibitem{Murdia:2019fax}
C.~Murdia, Y.~Nomura, P.~Rath and N.~Salzetta, {\it {Comments on holographic
  entanglement entropy in $TT$ deformed conformal field theories}},  {\em Phys.
  Rev. D} {\bf 100} (2019), no.~2 026011
  [\href{http://arXiv.org/abs/1904.04408}{{\tt 1904.04408}}].

\bibitem{Ota:2019yfe}
T.~Ota, {\it {Comments on holographic entanglements in cutoff AdS}},
  \href{http://arXiv.org/abs/1904.06930}{{\tt 1904.06930}}.

\bibitem{Banerjee:2019ewu}
A.~Banerjee, A.~Bhattacharyya and S.~Chakraborty, {\it {Entanglement Entropy
  for $TT$ deformed CFT in general dimensions}},  {\em Nucl. Phys. B} {\bf 948}
  (2019) 114775 [\href{http://arXiv.org/abs/1904.00716}{{\tt 1904.00716}}].

\bibitem{Banados:1998gg}
M.~Banados, {\it {Three-dimensional quantum geometry and black holes}},
  \href{http://arXiv.org/abs/hep-th/9901148}{{\tt hep-th/9901148}}.
%%CITATION = HEP-TH/9901148;%%

\bibitem{Ammon:2012wc}
M.~Ammon, M.~Gutperle, P.~Kraus and E.~Perlmutter, {\it {Black holes in three
  dimensional higher spin gravity: A review}},  {\em J.Phys.} {\bf A46} (2013)
  214001 [\href{http://arXiv.org/abs/1208.5182}{{\tt 1208.5182}}].
%%CITATION = ARXIV:1208.5182;%%

\bibitem{Witten:1998qj}
E.~Witten, {\it {Anti-de Sitter space and holography}},  {\em Adv. Theor. Math.
  Phys.} {\bf 2} (1998) 253--291
  [\href{http://arXiv.org/abs/hep-th/9802150}{{\tt hep-th/9802150}}].
%%CITATION = HEP-TH/9802150;%%

\bibitem{Miskovic:2006tm}
O.~Miskovic and R.~Olea, {\it {On boundary conditions in three-dimensional AdS
  gravity}},  {\em Phys. Lett.} {\bf B640} (2006) 101--107
  [\href{http://arXiv.org/abs/hep-th/0603092}{{\tt hep-th/0603092}}].
%%CITATION = HEP-TH/0603092;%%

\bibitem{deBoer:2013gz}
J.~de~Boer and J.~I. Jottar, {\it {Thermodynamics of higher spin black holes in
  $AdS_3$}},  {\em JHEP} {\bf 1401} (2014) 023
  [\href{http://arXiv.org/abs/1302.0816}{{\tt 1302.0816}}].
%%CITATION = ARXIV:1302.0816;%%

\bibitem{Kraus:2006wn}
P.~Kraus, {\it {Lectures on black holes and the AdS(3) / CFT(2)
  correspondence}},  {\em Lect.Notes Phys.} {\bf 755} (2008) 193--247
  [\href{http://arXiv.org/abs/hep-th/0609074}{{\tt hep-th/0609074}}].
%%CITATION = HEP-TH/0609074;%%

\bibitem{Gubser:2002vv}
S.~S. Gubser and I.~R. Klebanov, {\it {A Universal result on central charges in
  the presence of double trace deformations}},  {\em Nucl. Phys.} {\bf B656}
  (2003) 23--36 [\href{http://arXiv.org/abs/hep-th/0212138}{{\tt
  hep-th/0212138}}].
%%CITATION = HEP-TH/0212138;%%

\bibitem{Papadimitriou:2007sj}
I.~Papadimitriou, {\it {Multi-Trace Deformations in AdS/CFT: Exploring the
  Vacuum Structure of the Deformed CFT}},  {\em JHEP} {\bf 05} (2007) 075
  [\href{http://arXiv.org/abs/hep-th/0703152}{{\tt hep-th/0703152}}].
%%CITATION = HEP-TH/0703152;%%

\bibitem{Bzowski:2018pcy}
A.~Bzowski and M.~Guica, {\it {The holographic interpretation of $J \bar
  T$-deformed CFTs}},  {\em JHEP} {\bf 01} (2019) 198
  [\href{http://arXiv.org/abs/1803.09753}{{\tt 1803.09753}}].
%%CITATION = ARXIV:1803.09753;%%

\bibitem{Strobl:1999wv}
T.~Strobl, {\em {Gravity in two space-time dimensions}}.
\newblock PhD thesis, Aachen, Tech. Hochsch., 1999.
\newblock \href{http://arXiv.org/abs/hep-th/0011240}{{\tt hep-th/0011240}}.
%%CITATION = HEP-TH/0011240;%%

\bibitem{Zamolodchikov:2004ce}
A.~B. Zamolodchikov, {\it {Expectation value of composite field T anti-T in
  two-dimensional quantum field theory}},
  \href{http://arXiv.org/abs/hep-th/0401146}{{\tt hep-th/0401146}}.
%%CITATION = HEP-TH/0401146;%%

\bibitem{Guica:2019nzm}
M.~Guica and R.~Monten, {\it {$T\bar T$ and the mirage of a bulk cutoff}},
  \href{http://arXiv.org/abs/1906.11251}{{\tt 1906.11251}}.
%%CITATION = ARXIV:1906.11251;%%

\bibitem{Balasubramanian:2002zh}
V.~Balasubramanian, J.~de~Boer and D.~Minic, {\it {Notes on de Sitter space and
  holography}},  {\em Class. Quant. Grav.} {\bf 19} (2002) 5655--5700
  [\href{http://arXiv.org/abs/hep-th/0207245}{{\tt hep-th/0207245}}]. [Annals
  Phys.303,59(2003)].
%%CITATION = HEP-TH/0207245;%%

\bibitem{Kraus:2006nb}
P.~Kraus and F.~Larsen, {\it {Partition functions and elliptic genera from
  supergravity}},  {\em JHEP} {\bf 0701} (2007) 002
  [\href{http://arXiv.org/abs/hep-th/0607138}{{\tt hep-th/0607138}}].
%%CITATION = HEP-TH/0607138;%%

\bibitem{Carroll:1997ar}
S.~M. Carroll, {\it {Lecture notes on general relativity}},
  \href{http://arXiv.org/abs/gr-qc/9712019}{{\tt gr-qc/9712019}}.
%%CITATION = GR-QC/9712019;%%

\end{thebibliography}\endgroup

\end{document}